\documentclass[a4paper,aps,prd,10pt,preprintnumbers,twocolumn,superscriptaddress,nofootinbib,amsmath,amssymb]{revtex4-1}
\usepackage[dvips]{graphics}
\usepackage[utf8]{inputenc}
\usepackage[T1]{fontenc}
\usepackage{cmap}

\begin{document}
\title{Echoes of compact objects: new physics near the surface and matter at a distance}
\author{R. A. Konoplya}\email{konoplya\_roma@yahoo.com}
\affiliation{Institute of Physics and Research Centre of Theoretical Physics and Astrophysics, Faculty of Philosophy and Science, Silesian University in Opava, CZ-746 01 Opava, Czech Republic}
\affiliation{Peoples Friendship University of Russia (RUDN University), 6 Miklukho-Maklaya Street, Moscow 117198, Russian Federation}
\author{Z. Stuchlík}\email{zdenek.stuchlik@fpf.slu.cz}
\affiliation{Institute of Physics and Research Centre of Theoretical Physics and Astrophysics, Faculty of Philosophy and Science, Silesian University in Opava, CZ-746 01 Opava, Czech Republic}
\author{A. Zhidenko}\email{olexandr.zhydenko@ufabc.edu.br}
\affiliation{Institute of Physics and Research Centre of Theoretical Physics and Astrophysics, Faculty of Philosophy and Science, Silesian University in Opava, CZ-746 01 Opava, Czech Republic}
\affiliation{Centro de Matemática, Computação e Cognição (CMCC), Universidade Federal do ABC (UFABC),\\ Rua Abolição, CEP: 09210-180, Santo André, SP, Brazil}
\begin{abstract}
It is well known that a hypothetical compact object that looks like an Einsteinian (Schwarzschild or Kerr) black hole everywhere except a small region near its surface should have the ringdown profile predicted by the Einstein theory at early and intermediate times, but modified by the so-called \emph{echoes} at late times. A similar phenomenon appears when one considers an Einsteinian black hole and a shell of matter placed at some distance from it, so that astrophysical estimates could be made for the allowed mass of the black hole environment. While echoes for both systems have been extensively studied recently, no such analysis has been done for a system featuring phenomena simultaneously, that is, echoes due to new physics near the surface/event horizon and echoes due to matter at some distance from the black hole. Here, following \cite{Damour:2007ap,Cardoso:2016rao}, we consider a traversable wormhole obtained by identifying two Schwarzschild metrics with the same mass $M$ at the throat, which is near the Schwarzschild radius, and add a nonthin shell of matter at a distance. This allows us to understand how the echoes of the surface of the compact object are affected by the astrophysical environment at a distance. The straightforward calculations for the time-domain profiles of such a system support the expectations that if the echoes are observed, they should most probably be ascribed to some new physics near the event horizon rather than some ``environmental'' effect.
\end{abstract}
\pacs{04.50.Kd,04.70.Bw,04.30.-w,04.80.Cc}
\maketitle

\section{Introduction}

Recent observations of black holes in the gravitational \cite{TheLIGOScientific:2016src} and electromagnetic \cite{Goddi:2017pfy,Bambi:2015kza} spectra have provided the opportunity to test the strong-gravity regime via black holes. The data from the purely gravitational spectrum during the ringdown phase still allow for large deviations from Kerr geometry due to a huge uncertainty in the determination of the angular momentum and mass of the resultant black hole \cite{Konoplya:2016pmh,Yunes:2016jcc}. Nevertheless, further constraints on the parameters of alternative theories of gravity are expected \cite{Berti:2018vdi}. While there remains the possibility of an essentially non-Kerr black-hole geometry owing to a non-Einsteinian gravitational theory, there may also be a more subtle situation: the black hole is Einsteinian (given by the Schwarzschild or Kerr geometry) everywhere except for a tiny region near the event horizon. In this case, the quasinormal ringing of a black hole (or even of a more exotic compact object, such as a gravastar \cite{Visser:2003ge,Chirenti:2007mk} or wormhole \cite{Damour:2007ap}) mimics that of a Schwarzschild/Kerr black hole very well \cite{Konoplya:2016hmd}, except for possibly the very late period, which will be modified by the so-called \emph{echoes} \cite{Tsang:2018uie,Cardoso:2017cqb,Cardoso:2016oxy,Cardoso:2016rao}. The effective potential for the master equation describing perturbations of the Schwarzschild spacetime has the form of a potential barrier and the quasinormal modes are poles of the reflection coefficient for that barrier. The echoes appear owing to the second scattering from the other peak of the effective potential near the event horizon and have been extensively studied recently for various compact objects and gravitational theories \cite{Nakano:2017fvh,Testa:2018bzd,Wang:2018mlp,Bueno:2017hyj,Maselli:2017tfq}. The second peak appears in a number of different circumstances, such as a different equation of state and boundary conditions on the surface of the ultracompact object (see fig.~3 in \cite{Cardoso:2016rao}) or a cloud of matter near the surface/horizon.

At the same time, large astrophysical-scale black holes are not believed to be free from the influence of their environment, be it accreting disks, other companion compact objects, active galactic nuclei, or clouds of normal and/or phantom matter. This gave rise to the concept of a ``dirty black hole'' \cite{Visser:1992qh,Krauss:1996rg,Babichev:2004yx,Macedo:2015ikq}. The effective potential in the above cases can have an additional peak (for phantom matter) or gap (for normal matter) in the far region, i.~e., farther than the main ``Schwarzschild peak''. Therefore, it would be natural to expect echoes from scattering near the far peaks as well \cite{Barausse:2014tra}. In other words, once the  echoes are observed, it will be crucial to understand whether the effect should be ascribed to new physical effects near the surface of a compact object or to some (possibly unseen) matter some distance from it.

From the theoretical point of view, for any particular compact object the gravitational theory and model of the surrounding matter would give us various detailed answers to the above question of how to distinguish both types of echoes, but a general qualitative understanding of whether both types of echoes produce equivalent effects, or if one of them could suppress the other, should come first. So far, two types of papers have considered echoes. One type has been devoted to echoes from compact objects due to modifications to Schwarzschild/Kerr geometry near the horizon/surface \cite{Nakano:2017fvh,Testa:2018bzd,Wang:2018mlp,Bueno:2017hyj,Maselli:2017tfq,Barcelo:2017lnx,Carballo-Rubio:2018jzw} (beginning with \cite{Cardoso:2016rao}), while the other group of works considered echoes from the astrophysical environment by placing a massive shell at some distance from the Schwarzschild black hole  \cite{Barausse:2014tra}. There was considered a shell of matter that is not infinitely thin and has either positive or negative energy (thus representing either normal or phantom matter). It was shown that the thickness of the shell does not drastically change the estimates. This generalized the approach of \cite{Leung:1999rh} based on an infinitely thin shell. The astrophysical estimations made in \cite{Barausse:2014tra} showed that the deviation from Schwarzschild ringdown is relatively small unless the mass of the shell is large enough, so that for the majority of astrophysical factors the effect should be relatively small. Nevertheless, possible configurations of dark matter around black holes would leave some parametric freedom for echoes as well \cite{Barausse:2014tra}.

Here we shall simultaneously consider both factors that can lead to echoes: the modification of the Schwarzschild geometry near the surface, and a nonthin shell of matter some distance from it. This is a straightforward way to realize how echoes owing to new physics near the surface would be affected by matter at a distance. We shall consider a traversable wormhole obtained by identifying two Schwarzschild metrics \cite{Damour:2007ap,Cardoso:2016oxy} and add a nonthin shell of matter at some distance from its throat. The echoes of such a wormhole alone were studied in \cite{Cardoso:2016oxy}.

The paper is organized as follows. Sec.~\ref{sec:basic} gives essential information on the construction of our configuration: the traversable wormhole built with the help of an infinitely thin shell of exotic matter at the throat and another massive shell at a distance representing the astrophysical environment. Sec.~\ref{sec:timedomain} is devoted to the wave equation, boundary conditions, and time-domain integration method used for our analysis of the evolution of perturbations. Sec.~\ref{sec:echoes} discusses the influence of the distant shell on echoes induced by a modification of the Schwarzschild spacetime near the wormhole's throat. Finally, in Sec.~\ref{sec:conclusions} we summarize the obtained results.

\section{Traversable thin-shell wormhole with a massive shell at a distance}\label{sec:basic}

Following \cite{Cardoso:2016oxy}, we consider a Damour-Solodukhin wormhole \cite{Damour:2007ap} which is obtained by identifying two Schwarzschild metrics with the same mass $M=0.5$ described by the line element
\begin{equation}\label{metric}
ds^2=-f(r)dt^2+\frac{dr^2}{f(r)}+r^2(d\theta^2+\sin^2\theta d\phi^2)\,,
\end{equation}
with
$$f(r)\equiv1-\frac{2M}{r}\,.$$

The surgery at the throat $r_0>2M$ requires a thin shell of matter with surface density and surface pressure
\begin{equation}
\Sigma =-\frac{\sqrt{1-2M/r_0}}{2\pi r_0}\,, \quad p =\frac{1}{4\pi r_0} \frac{(1-M/r_0)}{\sqrt{1-2M/r_0}}\,,
\end{equation}
respectively. The weak energy condition is violated as $\Sigma<0$, while the strong and null energy conditions are fulfilled when the throat is within the photosphere $r_0<3M$.
Notice that in the limit $r_0\to2M$ the throat approaches the event horizon and the model splits into two separate Schwarzschild black holes.
As an illustration, we consider the same model as in \cite{Cardoso:2016oxy} and choose $r_0=1.0000005$.

\begin{figure}
\resizebox{ \linewidth}{!}{\includegraphics*{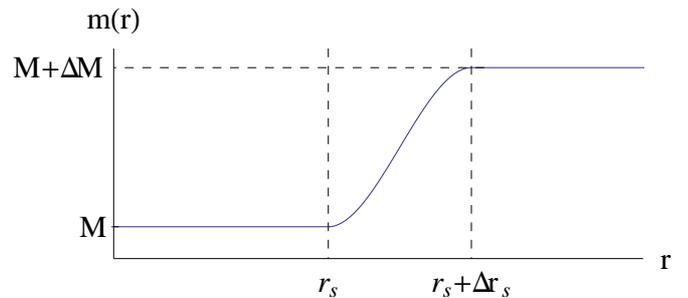}}
\caption{Choice of the mass function.}\label{fig:massfunc}
\end{figure}

We also add to the wormhole a shell of mass $\Delta M$ located between $r_s>r_0$ and $r_s+\Delta r_s$ such that the mass function is defined as
\begin{equation}\label{massfunc}
m(r)=\left\{
        \begin{array}{l}
          M,\phantom{\Delta} \qquad\qquad\qquad r<r_s\,; \\
          M+\Delta M\left(3-2\dfrac{r-r_s}{\Delta r_s}\right)\left(\dfrac{r-r_s}{\Delta r_s}\right)^2,  \\
          \phantom{\Delta~M,\!} \qquad\qquad\qquad r_s\leq r\leq  r_s+\Delta r_s\,;\\
          \Delta M, \qquad\qquad\qquad r_s+\Delta r_s<r\,;
        \end{array}
      \right.
\end{equation}
and
$$f(r)=1-\frac{2m(r)}{r}\,.$$

In this way, $m(r)$ and $m'(r)$ are continuous functions (see Fig.~\ref{fig:massfunc}). Here $\Delta M > 0$ ($\Delta M < 0$) corresponds to a positive (negative) energy density of matter.

\section{The wave equation and time-domain integration}\label{sec:timedomain}

For our qualitative consideration we only need to estimate the orders of the effects produced by both types of echoes; thus, it is sufficient to limit ourselves to a test field response to the initial perturbation. Even though perturbations of fields of other spins produce different quasinormal spectra, as a rule the dominant frequencies are of the same order \cite{reviews}. We shall consider the Klein-Gordon equation for a massless scalar field, which can be reduced to the wavelike form
\begin{equation}\label{eq:wavelike}
\left(\frac{\partial^2}{\partial t^2}-\frac{\partial^2}{\partial r_*^2}+V(r)\right)\Psi(t,r_*)=0\,,
\end{equation}
where $r_*$ is the tortoise coordinate in the observer's universe,
$$dr_*=\pm\frac{dr}{f(r)}\,.$$
Here the signs $\pm$ refer to the two different universes connected at the throat $r_0$, and the effective potential is given by
\begin{equation}
V(r) = f(r)\left(\frac{\ell(\ell+1)}{r^2}+\frac{f'(r)}{r}\right)\,,
\end{equation}
where $\ell=0,1,2,\ldots\,$ are the multipole numbers.

The whole space lies between two ``infinities'' connecting two distant regions or universes. Quasinormal modes of wormholes are solutions of the wave equation that satisfy the boundary conditions of having purely incoming waves at $-\infty$ and purely outgoing waves at $+\infty$ \cite{Konoplya:2005et,Konoplya:2010kv}.
This means that no waves coming from either left or right infinity are allowed. This way, the boundary conditions for a wormhole are essentially the same as those for a black hole and our conclusions are expected to be qualitatively the same if, for example, one considers a black hole that is modified near its event horizon, instead of a wormhole.

\begin{figure}
\resizebox{\linewidth}{!}{\includegraphics*{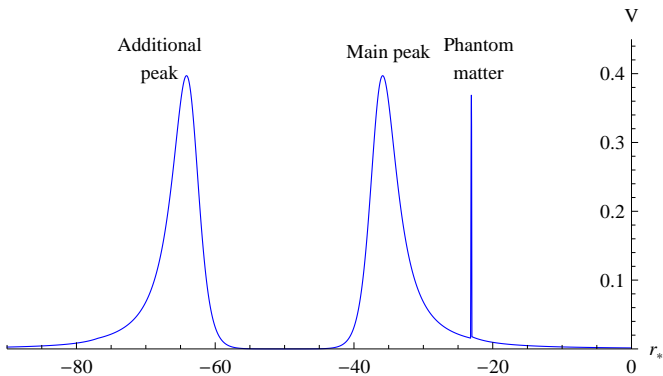}}
\caption{The effective potential ($\ell=1$) for a thin-shell wormhole of mass $M=0.5$ with a throat at $r_0=1.0000005$ with a shell at $r_s=11$ of size $\Delta r_s=0.2$ and mass $\Delta M=-2.5$. The ``main peak'' is from the Schwarzschild effective potential in the absence of either new physics at the surface or matter at a distance. The ``additional peak'' is due to the modification of the metric near the Schwarzschild radius.}\label{fig:DSWormholesPhantShellPot}
\end{figure}

A positive peak (fig.~\ref{fig:DSWormholesPhantShellPot}) corresponds to phantom matter ($\Delta M<0$). For the shell of positive mass the peak is replaced by a gap.

The ringdown phase for spherically symmetric perturbations ($\ell=0$) is relatively short. Although the results in this case are qualitatively similar, we will use higher $\ell$ to better illustrate the situation. Also, $\ell=0$ is not a dynamical degree of freedom for the gauge fields.

In order to produce the time-domain profiles, we integrate the wavelike equation (\ref{eq:wavelike}) rewritten in terms of the light-cone variables $u = t - r_*$ and $v = t + r_*$. The discretization scheme \cite{Gundlach:1993tp} has the form
\begin{eqnarray}\label{eq:scheme}
\Psi(N) &=& \Psi(W) + \Psi(E) -
\Psi(S)  \\
& &- \Delta^2\frac{V(W)\Psi(W) + V(E)\Psi(E)}{8} +
\mathcal{O}(\Delta^4)\ ,
\nonumber
\end{eqnarray}
where we have used the following definitions for the points: $N = (u + \Delta, v + \Delta)$, $W = (u + \Delta, v)$, $E = (u, v + \Delta)$ and $S = (u,v)$. The initial data are specified on the two null surfaces $u = u_{0}$ and $v = v_{0}$. This method was tested in a great number of papers (see, for example, recent works \cite{Konoplya:2018qov,Macedo:2018txb} and references therein) and showed good convergence and agreement with accurate calculations done using other approaches \cite{reviews}.

\section{Echoes}\label{sec:echoes}

\begin{figure}
\resizebox{\linewidth}{!}{\includegraphics*{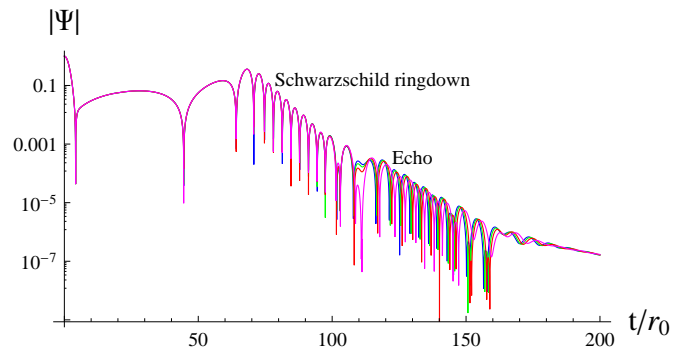}}
\caption{Time-domain profiles ($\ell=2$) at $r=30$ for a black hole ($r_0=2M=1$) with a heavy shell $\Delta M=0.25$ at $r_s=21$ for $\Delta r_s=0.1$ (blue), $\Delta r_s=0.5$ (green), $\Delta r_s=1$ (red), and $\Delta r_s=2$ (magenta). All profiles are qualitatively the same, with the echo stage starting after $\Delta t\approx40$ the quasinormal ringing stage. }\label{fig:varshellsize}
\end{figure}

As our configuration contains two shells, one is infinitely thin at the throat and the other is the distant shell representing matter; from here and on, when mentioning a ``shell'' we will mean the nonthin shell placed some distance from the wormhole's throat. We need to understand the dependence of the quasinormal ringing on the following characteristics of our configuration: the mass of the shell $\Delta M$, its position $r_s$, and thickness $\Delta r_{s}$. The latter characteristic determines the density of the shell and would intrinsically depend on the equation of state for the matter. Fortunately, we observe that the profiles of the quasinormal ringing depend very weakly on $\Delta r_{s}$ (see fig.~\ref{fig:varshellsize}), which supports the conclusions of \cite{Barausse:2014tra} on the shell configuration around a Schwarzschild black hole. As the quasinormal ringing does not depend much on the density and size of the shell, it also does not come as a surprise that the use of other types of mass functions \cite{Barausse:2014tra} does not qualitatively change the profiles.

The position of the shell does not affect the intensity of the echoes so much as it changes the time at which the echoes begin \cite{Mirbabayi:2018mdm}. For the shell representing matter around a black hole, the minimal distance (at which such a quasistationary configuration of matter is still justified) is determined by the innermost stable circular orbit (ISCO) at $r=3$. The most influential factor of the model is the mass of the shell.

The shell of normal or phantom mass can cause an echo-type signal. However, in order to produce a distinct second pulse the mass of the shell has to be of the same order as the black-hole mass. Apparently, the usual visible astrophysical environment such as stars, accreting disks, clouds of gas, etc.) should be many orders lighter than the black hole and is unlikely to produce a measurable distortion of the echoes from the surface. Nevertheless, this may not be so for dark matter/energy whose interaction with the black hole is largely unknown and may lead to qualitatively new phenomena \cite{Barausse:2014tra,Raidal:2018eoo,Babichev:2004yx}. Moreover, if such an enormous mass was indeed spread throughout a region around colliding black holes, it would lead to a significant change of the signal during the inspiral phase, which is much more sensitive to the external matter compared to the ringdown phase. This way, the signal would be distinctively non-Schwarzschild even at the stages that could be described by the post-Newtonian approximation.

\begin{figure}
\resizebox{\linewidth}{!}{\includegraphics*{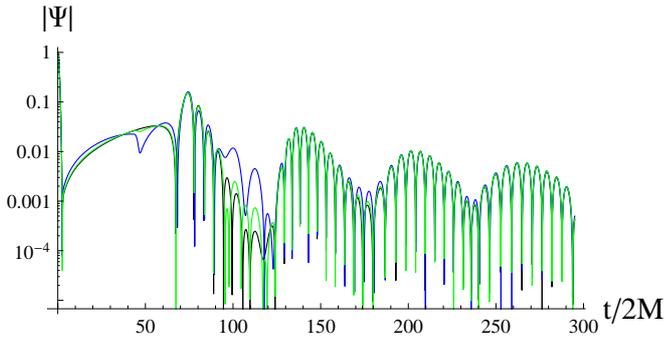}}
\caption{Profiles ($\ell=1$) for a thin-shell wormhole of mass $M=0.5$ with a throat at $r_0=1.0000005$. Black profile: ringdown without shell. Green profile: a shell at $r_s=11$ of size $\Delta r=1$ and mass $\Delta M=0.5$. Blue profile: a shell at $r_s=11$ of size $\Delta r_s=1$ and mass $\Delta M=2.5$. }\label{fig:DSWormholeshell}
\end{figure}

\begin{figure}
\resizebox{\linewidth}{!}{\includegraphics*{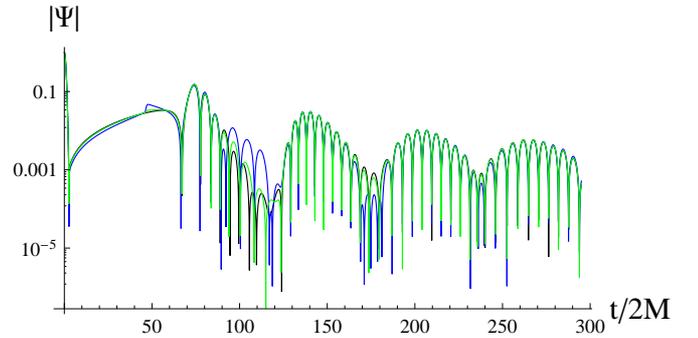}}
\caption{Profiles ($\ell=1$) for a thin-shell wormhole of mass $M=0.5$ with a throat at $r_0=1.0000005$. Black profile: ringdown without shell. Green profile: a shell at $r_s=11$ of size $\Delta r_s=0.2$ and mass $\Delta M=-0.5$. Blue profile: a shell at $r_s=11$ of size $\Delta r_s=0.2$ and mass $\Delta M=-2.5$. }\label{fig:DSWormholesPhantShellProfiles}
\end{figure}

Even a large mass of matter far from a black hole leads to a small correction to the effective potential, compared to the near-horizon geometry. For a Schwarzschild-like wormhole we observe that the echo signal due to reflection from the additional peak, formed by modifications near the surface/throat (see fig.~\ref{fig:DSWormholesPhantShellPot}), has a larger amplitude and dominates the echo due to the shell of matter at a distance (see figs.~\ref{fig:DSWormholeshell}~and~\ref{fig:DSWormholesPhantShellProfiles}).

If instead of the Schwarzschild-like wormhole we considered a pure Schwarzschild black hole, the power-law asymptotic tails would dominate in the signal at late times. Once one adds a shell of massive matter at a distance, the situation changes drastically: asymptotic tails appear at even later times, while immediately after the period of quasinormal oscillations  significant echoes from a distant massive shell appear instead. Thus, a massive shell at a distance could be distinguished from the purely Schwarzschild evolution of perturbations. This is not so when we have new physics near the surface of a compact object, such as a wormhole. In this case the strong echoes of the surface dominate the echoes of the distant shell and only an extraordinarily large mass  located sufficiently close to the wormhole would lead to relatively small but noticeable changes in the main echoes of the surface.

\section{Conclusions}\label{sec:conclusions}

Recently there have been many discussions about the phenomenon of echoes, which are deviations of the quasinormal ringing from its General Relativity profiles at sufficiently late times. This phenomenon takes place in two different situations: when there is a modification of the black-hole geometry only in a small region near its horizon or surface, or when some distribution of matter exists at a distance from the compact object. Here we have considered the traversable Schwarzschild-like wormhole of \cite{Cardoso:2016rao} and added a massive nonthin shell of matter which models the possible astrophysical environment of the compact object. We have shown that a distant shell, whose mass is much smaller than the wormhole, is unlikely to produce a measurable effect on echoes from the throat. The shell representing the surrounding astrophysical environment must be extraordinarily heavy (comparable to the mass of the compact object) to produce a noticeable effect on the ``main'' echoes. Such large masses are normally not expected for the usual visible astrophysical environment of compact objects, and if present they would drastically change the inspiral phase. Thus, we argue that \emph{if echoes are observed after the purely Einsteinian inspiral, merger, and early ringdown phases, then such echoes must be ascribed to potentially new physics near the surface of the compact object, rather than to any astrophysical environment.} It is also worth mentioning that even when the mass of the shell is of the same order or larger than the mass of the wormhole, the echoes remain unaffected  at sufficiently late times.

\acknowledgments{
R. K. acknowledges support of the International Mobility Project CZ.02.2.69\slash0.0\slash0.0\slash16\_027\slash0008521 and hospitality of the Silesian University in Opava. Z.~S.~acknowledges the Albert Einstein Centre for Gravitation and Astrophysics supported under the Czech Science Foundation (Grant No. 14-37086G).
A. Z. acknowledges support of the Research Centre for Theoretical Physics and Astrophysics, Faculty of Philosophy and Science of Silesian University at Opava.}

\end{document}